\documentclass{elsarticle}
\usepackage{hyperref}
\usepackage[margin=1in]{geometry}
\journal{Physica B}

\begin{document}

\begin{frontmatter}

\title{Inertial effects in systems with magnetic charge}

\author{N.P. Armitage}
\address{The Institute for Quantum Matter and the Department of Physics and Astronomy, The Johns Hopkins University, Baltimore, Maryland 21218, USA}
\ead{npa@pha.jhu.edu}

\begin{abstract}
This short article sets out some of the basic considerations that go into detecting the mass of quasiparticles with effective magnetic charge in solids.  Effective magnetic charges may be appear as defects in particular magnetic textures. A magnetic monopole is a defect in this texture and as such these are not monopoles in the actual magnetic field $\mathbf{B}$, but instead in the auxiliary field $\mathbf{H}$.   They may have particular properties expected for such quasiparticles such as magnetic charge and mass.  This effective mass may -- in principle -- be detected in the same fashion that the mass is detected of other particles classically e.g. through their inertial response to time-dependent electromagnetic fields.  I discuss this physics in the context of the ``simple" case of the quantum spin ices, but aspects are broadly applicable.   Based on extensions to Ryzkhin's model for classical spin ice, a hydrodynamic formulation can be given that takes into account inertial and entropic forces.  Ultimately, a form for the susceptibility is obtained that is equivalent to the Rocard equation, which is a classic form used to account for inertial effects in the context of Debye-like relaxation.

\end{abstract}


\end{frontmatter}


\section{Introduction}

The subject of isolated magnetic charges (e.g. monopoles) is a perennial one in modern physics.  Dirac argued that the existence of single magnetic monopole anywhere in the universe would account for the quantization of the electric charge \cite{Dirac1,Dirac2}.   And in grand unification theories magnetically charged particles appear naturally \cite{GUT}.  However, as is learned by all beginning physics students, no definitive signatures of fundamental magnetic charges have been found\footnote{On Valentine's day in 1982 an almost perfect signal of one Dirac unit of magnetic charge passed through a superconducting loop detector \cite{ValentineMonopole}.   Although not for lack of trying, the results of this Valentine's Day monopole ``massacre" were never reproduced.} and one can safely write the magnetic Gauss's Law as $\nabla \cdot \mathbf{B} = 0$.   However there are material systems where effective magnetic charge exists.   Generally speaking these magnetic charges can be stabilized because of some particular magnetic texture enforced in the low energy states.   The monopole is a defect in this texture and therefore these are not monopoles in the actual magnetic field $\mathbf{B}$, but instead in the auxiliary field  $\mathbf{H}$.   Nevertheless, in many circumstances they possess properties that one would expect for fundamental monopoles.   They have magnetic charge and may be manipulated in a fashion consistent with their particle nature.   And in many situations they can exhibit effects consistent with inertia and mass.

One of the simplest cases where such monopoles may arise is as domain walls in quasi-1D ferromagnetic nanowires made of classical soft ferromagnets.  The domain wall carries magnetic charge and can be manipulated by external fields.    In classical magnets as such, the magnetic texture encompasses many spins that give effective internal degrees of freedom that can store energy when they are in motion, which can give inertia and an effective mass e.g. effective hard modes of the system can impart inertia to soft modes when the hard modes are integrated out \cite{Oleg15a}.  D\"oring was the first to show that inertia can arise in this fashion for Bloch domain walls \cite{Doring}.  Domain walls can be moved by a $z$-axis applied homogeneous magnetic field.   The applied field exerts a torque on precessing spins in the domain wall, which pulls the spins out of the plane perpendicular to the direction of motion of the wall.   An internal field perpendicular to the wall is created that causes an additional component to the precession such that spin finally becomes parallel with the spins in a domain.   This distortion of the domain wall structure causes an increase of the domain wall energy that goes as the square of its velocity.

Inertia is the capacity of an object to resist changes in its velocity.   By evaluating the functional derivative of the kinetic energy with respect to position one can straightforwardly show that an object whose kinetic energy scales as $\frac{1}{2} m \textbf{v}^2 $ will have a contribution to its equation of motion that goes like $m \textbf{v}'$.   Hence the D\"oring mechanism can be associated with an effective inertia.   Effects of this kind were seen long ago in ferrite \cite{Rado50}.   More recently, Saitoh et al. \cite{Saitoh04} used an AC current to induce the resonant motion of a domain wall in Ni$_{81}$Fe$_{19}$.  Here, the effective magnetic charge is confined to a region between the two accessible ferromagnetic ground states.  The domain wall mass and magnetic charge are set by non-universal material parameters.  Saitoh et al. determine the mass of a domain wall of approximately $7.2 \times 10^7 m_e$, which is roughly consistent with the D\"oring mass prediction.  It has been shown recently that a related mechanism gives effective mass to skyrmions in the related problem of skyrmion dynamics in thin films, however with an additional enhancement due to their topological nature and deformation in shape when in motion \cite{Makhfudz,Buttner}.

In quantum magnets, where the fundamental excitations are flipped spins or the boundaries between spins, considerations and the generation of inertia are different.  Magnetic charges are again quite natural in 1D ferromagnetic spin chains, such as CoNb$_2$O$_6$ \cite{Coldea,Morris}.  One envisions the elementary spin flip (induced by say a neutron or photon) in the ferromagnetic ground state ``fractionalizing" into two domain wall ``kinks" that may propagate separately.   Each domain wall may again be considered a magnetic monopole.  Here inertial effects arise through intrinsically quantum mechanical spin flips terms as flipping a spin at the domain wall boundary is equivalent to domain wall tunneling.   The energy of a such a domain wall will generally go as its momentum squared.

In higher dimensional quantum magnets, more specific considerations must be given, but again defects in the certain ground state magnetic textures in quantum magnets can behave as effective monopoles.  One such system that has attracted a lot of attention in this regard are the spin ices that may occur in both 2D \cite{2DSpin} and 3D \cite{Ryzhkin05,Castelnovo08}.  3D spin ices may be realized in magnetic pyrochlore oxides, in which magnetic ions sit at the vertices of corner sharing tetrahedra of the pyrochlore lattice \cite{Gardner10}.  In classical spin ice materials, such as Dy$_2$Ti$_2$O$_7$ and Ho$_2$Ti$_2$O$_7$, large classical spins are forced by strong crystal field anisotropy in the local $<$111$>$ direction, with a primarily ferromagnetic Ising spin interaction. The resulting ground state is believed to obey the a version of the Bernal-Fowler ``ice rules", where each tetrahedron adopts the so-called ``two-in, two-out" configuration \cite{Bramwell01}.  This configuration is equivalent to that of proton disorder in water ice (two-close, two-far) and hence the classical spin ices have been found to similarly posses an extensive low temperature residual entropy \cite{Ramirez99a}.  Flipping a single spin (e.g, a dipole excitation) in the spin ice creates two violations of the spin-ice rule (``three-in, one-out" and ``one-in, three-out") in neighboring tetrahedra.   These defects are each half a dipole excitation and -- putting aside the short distance physics -- function as magnetic monopoles.  Through subsequent spin flips, the monopoles can be  separated without incurring any further violations of the ice rules giving the possibility of deconfined magnetic charges \cite{Ryzhkin05, Castelnovo08}.

The slow dynamics of the monopole motion in the classical spin ices are still a subject of investigation, but they are believed to be driven by strong fluctuating transverse component of the dipolar field arising from thermally fluctuating neighboring spins \cite{Ehlers03,Morris09}.  In contrast to these materials, there are systems that are believed to have finite transverse terms in the spin Hamiltonian (like the 1D quantum Ising system discussed above) through which monopole dynamics may become inherently quantum and change the situation dramatically.  One material that has been proposed to be such a ``quantum spin ice" is Yb$_2$Ti$_2$O$_7$.  The crystal field structure of this material allows the well isolated ground state Kramers doublet to be treated as an effective spin 1/2 moment in the low energy sector \citep{Hodges01,Malkin04}. The exchange parameters of Yb$_2$Ti$_2$O$_7$ are still a matter of some debate
 \cite{Ross11a,Applegate12, DOrtenzio13,Pan14,Robert15a,Thompson17a,Scheie17a} and there is controversy whether or not $J_{zz}$ is in fact larger than the transverse exchanges to put it in the quantum spin ice regime.   In this manuscript, I will ignore the (otherwise) important fact of whether or not a precise physical realization of quantum spin ice occurs and consider a model of the dynamics of some idealized version of quantum spin ice where the Ising terms are the largest, but there is still substantial transverse exchanges.
 
 In the present discussion, by quantum spin ice I mean not a \textit{state of matter} per se, but a \textit{phenomenology}.   I consider the quantum spin ice state to be a material that has local spin correlations that manifest the spin ice rules, but where defects in the spin ice configuration (e.g. monopoles) can tunnel.   For our purposes, the microscopic origin of the tunneling is unimportant, it is only essential that the timescales associated with it are shorter than time scales associated with incoherent thermally assisted spin flip processes.   For the quantum spin ices, it is assumed that it can arise from some manner of transverse exchange.    We will capture the basic phenomenology through a classical model of diffusing charges subject to inertial and entropic forces.  For the considerations here, there is no more contradiction inherent in using a classical model to describe a \textit{quantum} spin ice than there is to use a classical model to describe the conduction of thermally excited charge in a semiconductor where classical inertia also arises through inherently quantum tunneling.
 
The explicit motivation for my interest in all of this is recent time-domain THz experiments on Yb$_2$Ti$_2$O$_7$ \cite{Pan14,Pan15}.   Time-domain THz is a quasi-optical technique in which approximately \textit{ps} long pulses are transmitted through a material system and detected in a coherent fashion.   In the present cases of interest the time-varying magnetic field couples to the magnetic dipole degrees of freedom and allows a measure of the complex magnetic susceptibility.   The capacity to measure the complex magnetic susceptibility in the THz range is a unique capability of such experiments and allows for instance the detection of a sign change as a function of frequency in the real component of the THz susceptibility in Yb$_2$Ti$_2$O$_7$.  Putting aside the question of whether Yb$_2$Ti$_2$O$_7$ is a quantum spin ice or not, it is important to note that a sign change as such cannot be modeled with any classical equation of motion that does not incorporate an inertial term e.g. a term that depends on the second time derivative of the coordinate of interest.

 \section{Considerations for the AC Susceptibility}

First, let us consider the case of classical spin ices.   Although an analogy to driving electric charge with an electric field may be useful to get some intuition, the case of monopoles living in the particular vacuum of spin ice presents some complications. For starters, a monopole spin ice system cannot sustain a DC current. This is not just the ``difficult" technical issue associated with the injection of monopole charge from the ordinary physical vacuum into the spin ice vacuum\footnote{This is not a real constraint on even the Drude-like transport of electrons in ordinary metals. For instance the reflectivity of a piece of copper is independent of whether or not the copper is connected to an electrical ground.}, but a fundamental constraints on monopole transport in spin ice.  As monopole/anti-monopole pairs propagate away from each under the influence of an applied magnetic field they leave a ``Dirac string" of flipped spins behind that due to the constraints of the ice rules impedes a monopole charge of the same polarity to follow its path.  In a remarkable paper which was also the first to make the analogy between spin ice in pyrochlores and water ice,  Ryzhkin \cite{Ryzhkin05}  derived an expression for the magnetic relaxation of monopole motion in classical spin ice.  Here an analogy was made with previous work done on proton disorder and charge relaxation in water ice.  Starting from the formalism for the entropy production in an irreversible process he derived the monopole current to be

\begin{equation}
\mathbf{J}_i = \mu n_{i,m}(q_{i,m} \mathbf{H} - \eta_i \Phi \mathbf{ \Omega})
\label{Current}
\end{equation}

\noindent where $i$ refers to positive and negative monopole charge species, $\mu$ is the mobility of the monopoles, $n_{i,m}$ is their density, and $\eta_{i} = \pm 1$.  $q_{i,m}$ is the monopole effective charge which in the Ising limit is set by the condition that the spin moment $\mu_m $ equals $ q_m a /2 $ where $a$ is the distance between tetrahedral centers.

Ryzhkin's expression differs from a simple model of the transport of free charge (e.g. Ohm's law) by the inclusion of the 2nd term which includes a dependence on the configuration vector  $\mathbf{\Omega}$, which is related to the magnetization as  $\mathbf{\Omega} = \mathbf{M}/ q_m $.  A monopole current arises due to spin-flips in the current direction and hence corresponds to an increasing magnetization.  Finite magnetization reduces entropy and gives a thermodynamic force which opposes the current.   This reaction force originates in the configurational entropy of the spin ice vacuum and even in the absence of sample boundaries prevents a true dc current.  $\Phi$ is a proportionality constant derived to be $\frac{8}{\sqrt{3}} a k_B T$ for water ice models and is relevant here \cite{Ryzhkin97}.   Magnetic relaxation proceeds through defect motion with the configuration vector related to the monopole current history as 

\begin{equation}
\mathbf{ \Omega}(t) - \mathbf{ \Omega}(0)  = \int_0^t dt' (  \mathbf{ J}_+ - \mathbf{ J}_-).  
\label{Configuration}
\end{equation}

Taking the Fourier transform of Eq. \ref{Configuration} and substituting it into Eq. \ref{Current} gives a ``Debye-like" form for the susceptibility   ($\chi_m = \chi_m' + i\chi_m''$)

\begin{equation}
\chi_m(\omega) = \frac{q_m^2/ \Phi} { 1 - i \omega \tau}
\label{Debye}
\end{equation} 

\noindent with a relaxation time $\tau = \frac{q_m}{ n_m  \mu \Phi } $ in which $n_m$ is the total monopole density $ n_{+m} + n_{-m}$.  In general, the leading dependence of the susceptibility in powers of $\frac{1}{-i \omega } $ indicates the dominant term in the equations of motion.   Accordingly Eq. \ref{Debye} is characterized by a $\frac{1}{-i \omega } $ fall-off at high frequencies that reflects the dominant dissipative response in this model.   At low frequencies $\chi_m$ is constant and real (e.g. the zeroth power of  $\frac{1}{-i \omega } $) demonstrating that in the dc limit the dominant effect originates in the reaction force.  Note that in the Ryzhkin model both the real and imaginary components of Eq. \ref{Debye} are positive for all frequencies e.g. there is change of sign in $\chi'_m$.

Although Debye-like relaxation has been found to be an excellent description of the relaxation processes in classical spin ice (Ho$_2$Ti$_2$O$_7$ and Dy$_2$Ti$_2$O$_7$) at the low (kHz) frequencies relevant to those materials,  it is not a fully mathematically consistent response function as it does not fall off fast enough at high $\omega$ to satisfy various sum rules for instance, the zero temperature first moment sum rule  \cite{Hohenberg74,Zaliznyak05}, which is

\begin{equation}
\int_0^\infty d \omega \;  \omega  \chi_{ m
 \alpha\alpha}''(\mathbf{q}, \omega)  =  \frac{1}{2 \pi N \hbar^2} \langle [S_\mathbf{q}^\alpha, [{\cal H}, S_\mathbf{-q}^\alpha]] \rangle,
 \end{equation} 
 where  $S_\mathbf{q}^\alpha = \sum_j e^{-i \mathbf{q} \cdot \mathbf{R}_j  } S_{j}^\alpha$ is the Fourier transform of lattice spin operators and  ${\cal H}$ is a spin hamiltonian.   The value of this sum rule depends on aspects of the exchange interaction, but as it is a double commutator of the spin Hamiltonian with spin operators, will be finite.

A similar violation of a sum rule is a well-known defect of the Debye model in its application to electronic charge relaxation in insulators \cite{Onodera93}.   In general, the first moment of the $electric$ susceptibility (proportional to the conductivity) $must$ obey the sum rule $\int_0^\infty d \omega \;   \omega  \chi_e''  = \frac{\pi}{2} \frac{n_e e^2}{m_e}$ where $n_e$ is the total charge density and $m_e$ the electron mass.   In a fashion similar to the Ryzhkin model, the Debye model does not fall off fast enough at high $\omega$ to satisfy the sum rule.  The addition of an inertial term to the Debye model in the classical equations of motion gives a $\chi_m$ a high frequency asymptote that goes like $\frac{1}{- \omega^2}$ which insures the first moment's integrability \cite{Onodera93}.  Although at low frequencies in dissipative media, inertia can be neglected, at high enough frequencies it is essential for satisfying the sum rule.   In optical studies of electronic systems a restricted low energy sum rule is frequently applied, where the high frequency integration limit is taken to be finite, but with $n_e$ replaced by the electron density in the low energy sector, and $m_e$ a mass which is renormalized by band and/or interaction effects.  Therefore, for spin ice, irrespective of their exact origin, effects beyond Ryzhkin's treatment must be relevant at high frequency to satisfy the first moment sum rule.  In a similar fashion to electric charge relaxation in insulators, the Ryzhkin expression Eq. \ref{Debye} can be extended by including a phenomenological inertial term to Eq. \ref{Current}.  Quite generally, it will have the form of the final term in the expression

\begin{equation}
\mathbf{J}_i = \mu n_{i,m}(q_{i,m} \mathbf{H} - \eta_i \Phi \mathbf{ \Omega}) - \mathbf{\dot{J}}_i / \gamma 
\label{CurrentInertia}
\end{equation}

\noindent where $\gamma$ is the rate of current relaxation.   

With standard definitions the mobility is $ \mu = q_m/ \gamma m  $.  In what follows, $m$ is an effective inertial mass which arises in the low energy sector through non-Ising exchanges that lead to monopole tunneling.  Solving in the same fashion as in the  Ryzhkin case gives a classical equation of motion, the terms of which are immediately familiar as

\begin{equation}
\mathbf{\ddot{M}} + \gamma  \mathbf{\dot{M}} + \frac{n_m \Phi}{m} \mathbf{M}  = \frac{n_m q_m^2}{m}  \mathbf{H}.
\label{InertiaEquationMotion}
\end{equation}

Solving for the susceptibility we have 

\begin{equation}
\chi_m(\omega) = \frac{q_m^2 / \Phi }{ 1  - i \omega \gamma  m / n_m \Phi  - \omega^2  m / n_m \Phi  },
\label{SusceptibilityInertia}
\end{equation}

\noindent which in terms of $\tau$ (and $\gamma$) reads

\begin{equation}
\chi_m(\omega) = \frac{q_m^2 / \Phi }{ 1 - i \omega \tau   - \omega^2  \tau/ \gamma }.
\label{SusceptibilityInertia2}
\end{equation}

With substitutions $\omega_0^2 =  \frac{n_m \Phi}{m}  $ and $\omega_p^2 =  \frac{n_m q_m^2}{m}$  Eq. \ref{SusceptibilityInertia} is equivalent to the Drude-Lorentz equations, which may describe the response of excitations in a classical $electric$ charge oscillator. 

\begin{equation}
\chi_m(\omega) = \frac{\omega_p^2  }{ \omega_0^2     - \omega^2  - i \omega \gamma  }.
\label{SusceptibilityInertia3}
\end{equation}

\noindent Note  Eq. \ref{SusceptibilityInertia2} exhibits a sign change in the expression for $\chi'$ that is absent in  Ryzhkin's expression.   As discussed below, this sign change is diagnostic of inertial effects.  In the limit where $\gamma \rightarrow \infty$ Eq. \ref{SusceptibilityInertia2} recovers Ryzhkin's expression for Debye-like relaxation.   In this limit the zero crossing in $\chi'$  moves to infinity.

From Eq. \ref{SusceptibilityInertia2} for the susceptibility and in analogy with charge conductivity we can define a magnetic monopole conductivity $\kappa(\omega) = - i \omega \chi_m(\omega) = \kappa' + i \kappa ''$\footnote{It is important to note that this monopole conductivity $\kappa$  is not the same as the spin conductivity which is typically defined as the response of a spin current to the magnetic field \textit{gradient} in the small momentum limit \cite{Scalapino93a,Alvarez02a}.}.   Due to mapping of our expressions Eqs. \ref{SusceptibilityInertia} and \ref{SusceptibilityInertia2} to the Drude-Lorentz model, it is clear that the magnetic conductivity must obey a low energy sum rule \cite{WootenBook} which is 

\begin{equation}
\int_0^{\omega^*} d \omega \;  \kappa'(\omega) =   \frac{\pi}{2} \frac{n_m q_m^2}{m} .
\label{KappaSumRule}
\end{equation}
Here $\omega^*$ is a energy cut-off that must be smaller than a scale on the order of the Ising exchange parameter  $J_{zz}$, but much larger than $\gamma$.   Note that this is a different consideration to the conventional first moment sum rule for spin systems given above as $\kappa = - i \omega \chi_m$.  It would be interesting to evaluate the double commutator expression mentioned above for a model Hamiltonian and find an analogous expression sensitive to the magnetic charge. In the context of the present model, from fits to the monopole conductivity, in principle one can extract a spectral weight that is related to a monopole plasma frequency and more fundamental monopole parameters as $\omega_p^2 = n_m q_m^2/m_{m}$.   With knowledge of the monopole density $ n_m $, analysis of the fits to the susceptibility of the extended Ryzhkin model allow one to extract an effective mass $m_{m}$ of the monopoles.

\begin{figure*}
\includegraphics[trim = 0 5 5 5,width=7cm]{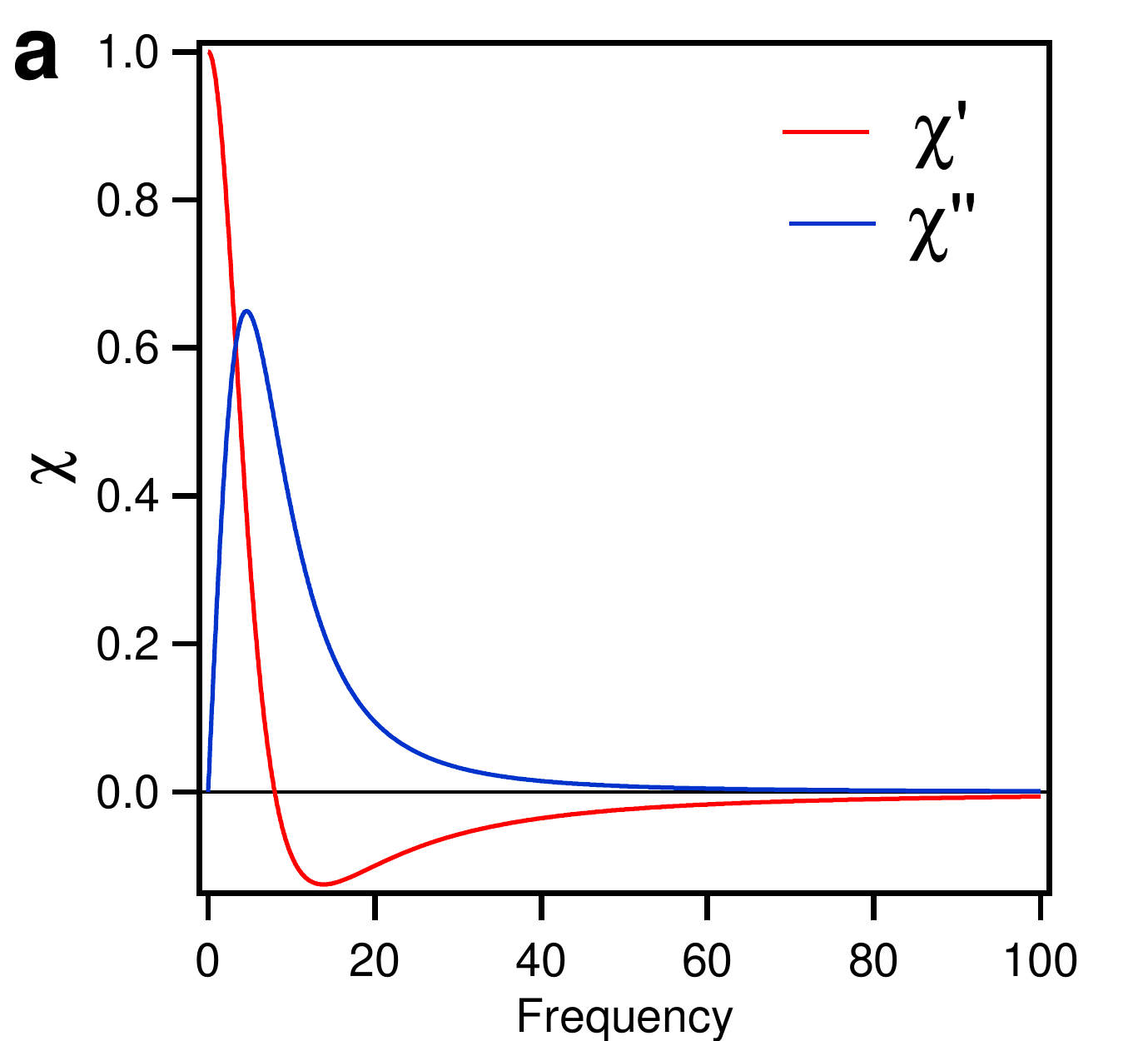}
\includegraphics[trim = 0 5 5 5,width=7cm]{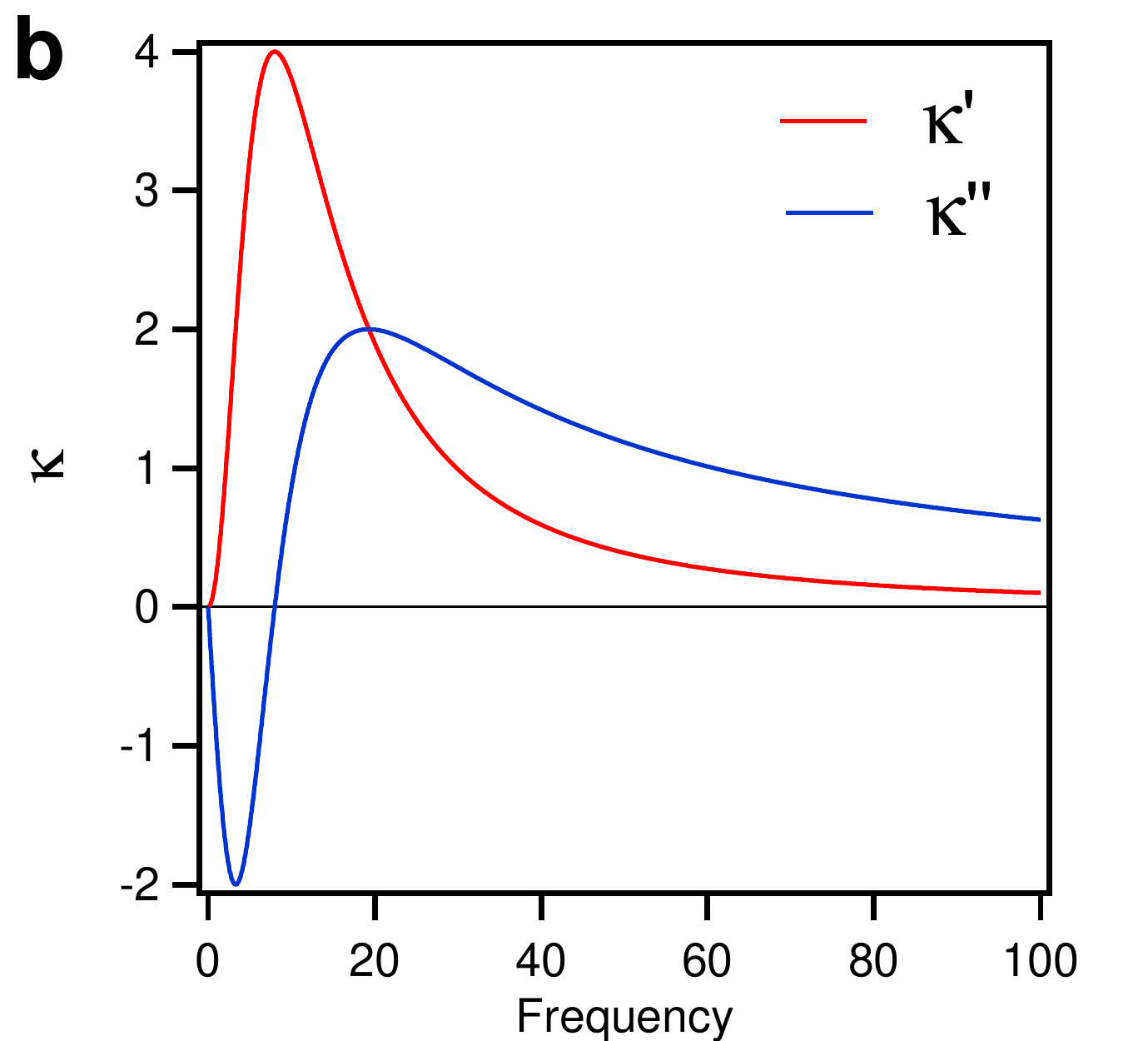}
\includegraphics[trim = 0 5 5 5,width=7cm]{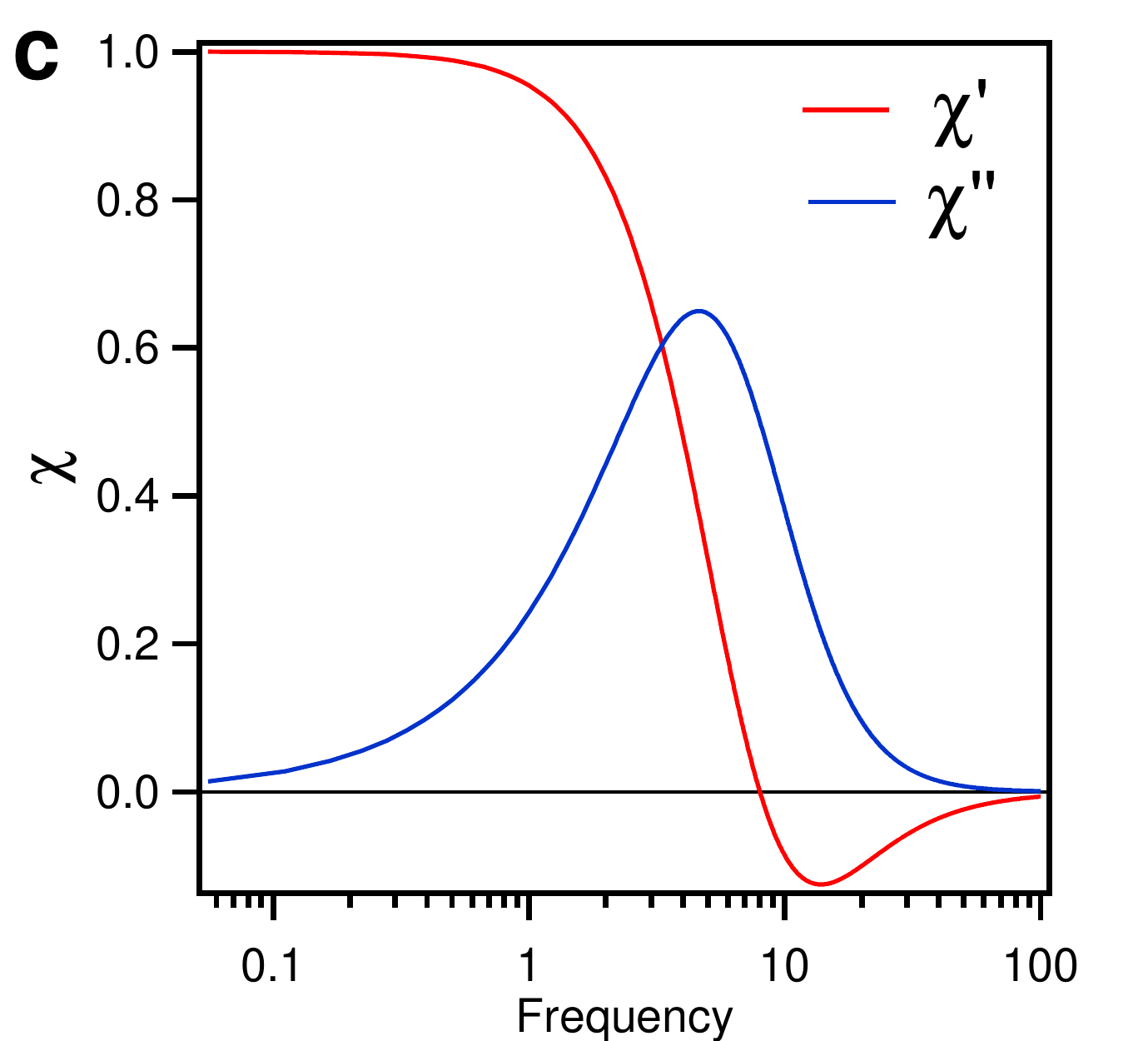}
\includegraphics[trim = 0 5 5 5,width=7cm]{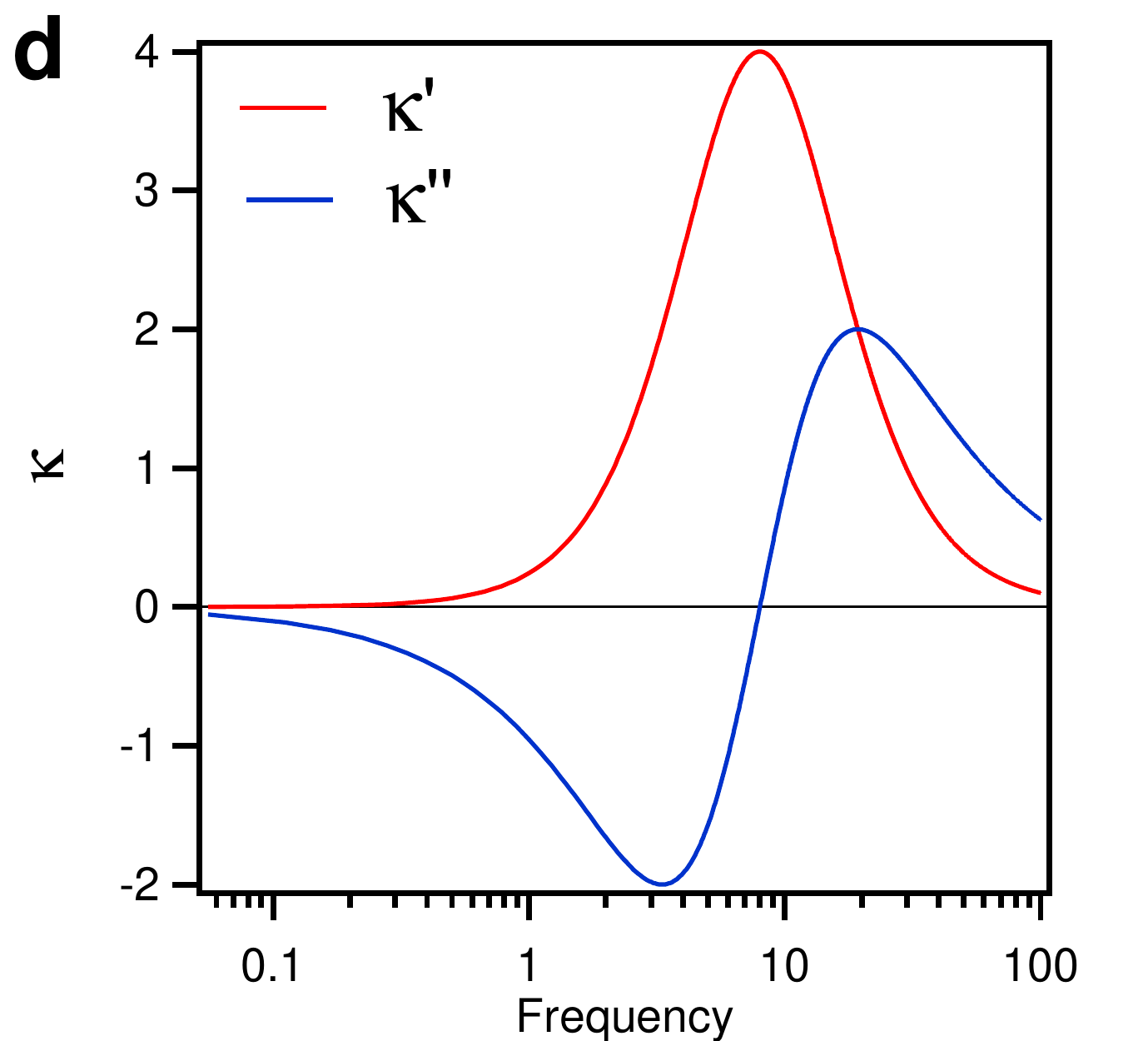}
\includegraphics[trim = 0 5 5 5,width=7cm]{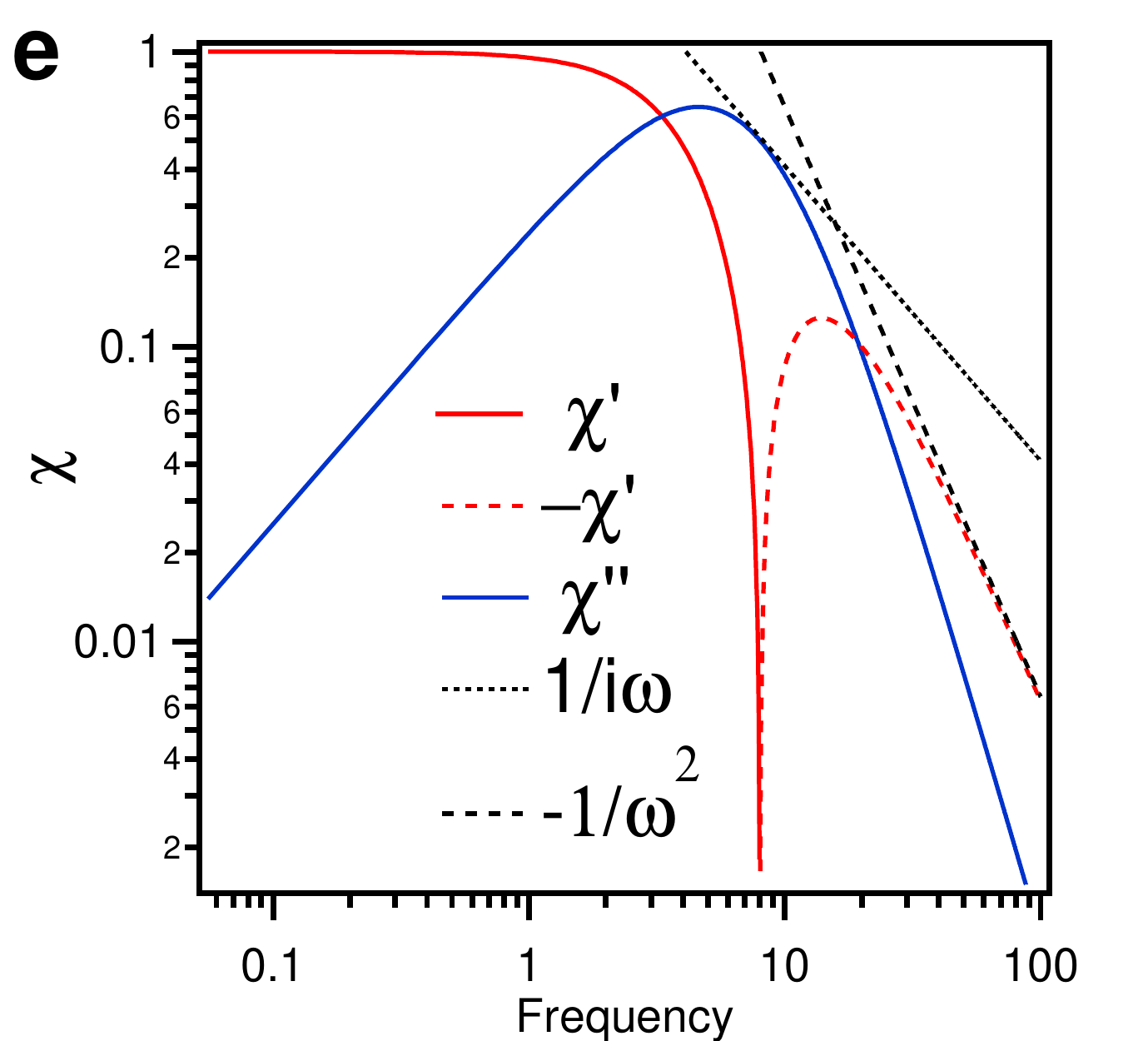}
\qquad \qquad
\includegraphics[trim = 0 5 5 5,width=7cm]{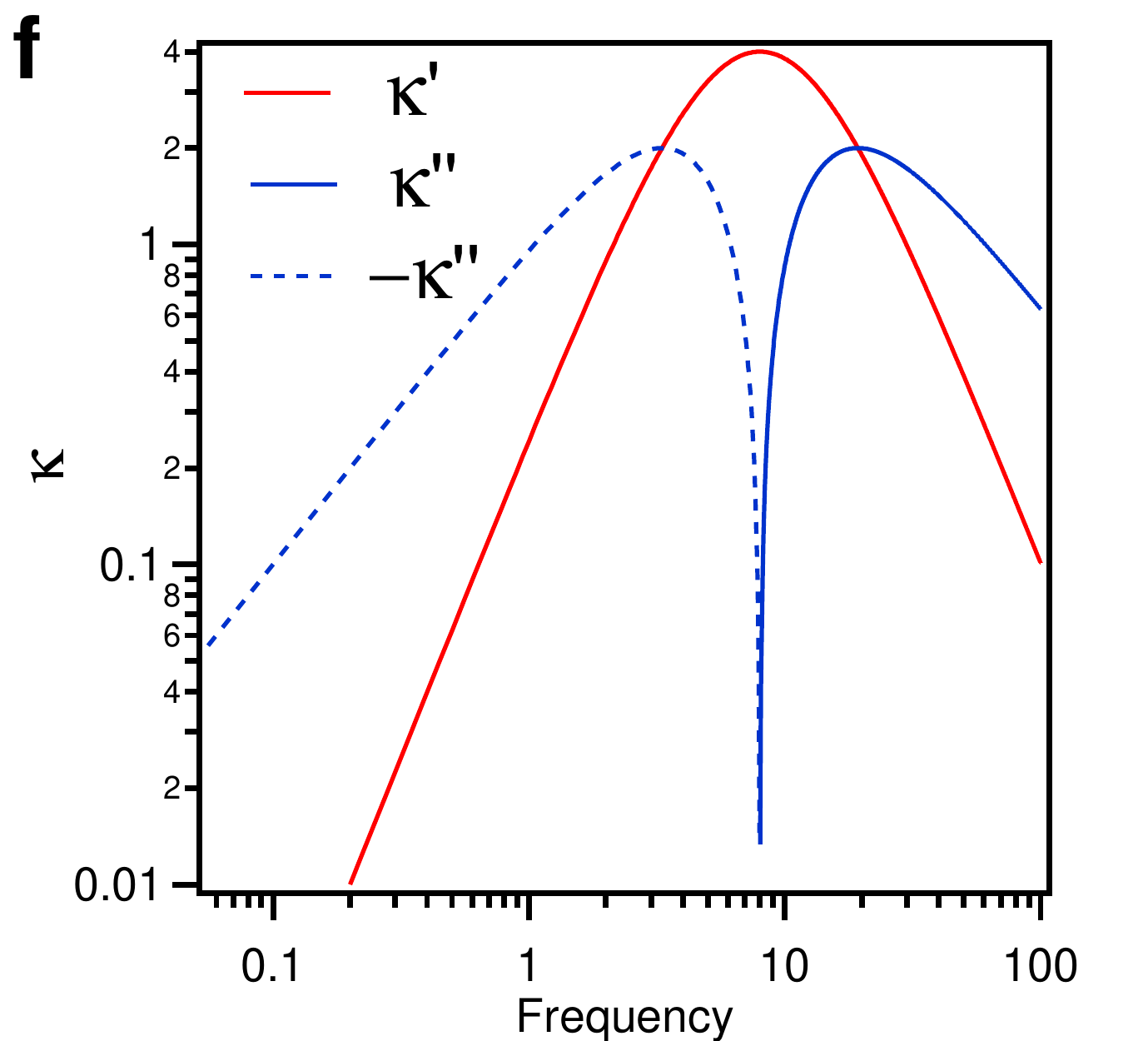}
\label{SI4} 
\caption{Plots of the susceptibility $\chi_m$ and the magnetic conductivity $\kappa$ with example parameters $1/ \tau = 4$ and $\gamma = 16$ on both linear, log, and log-log scales.}
\end{figure*}

\section{Simulations}

In Fig. 1, I plot the results of function Eq. \ref{SusceptibilityInertia2} for both susceptibility $\chi_m$ and the magnetic conductivity $\kappa$ for representative values of $\tau$ and $\gamma$ ($1/ \tau = 4$ and $\gamma = 16$) on both linear, log, and log-log scales.  These plots highlight the usefulness of analyzing both $\chi_m$ and $\kappa$.  The dissipative component of the susceptibility $\chi''$ peaks near $1/\tau$, which is the typical Debye-like relaxation behavior.  In $\chi_m$ the scale of $\gamma$ appears only as a subtle change in the high frequency power law, which manifests itself as a change of slope on the log-log plot.   In Fig. 2, I show a direct comparison of $\chi$ calculated both without and with the inertial term (e.g. Ryzhkin and extended Ryzhkin model).   One can see that the differences in $\chi"$ can be very subtle.  An experimentalist doing typical low frequency susceptibility or neutron scattering experiments (which measure only $\chi''$)  would be likely to be completely unaware of the high frequency scale where inertial effects become relevant as they manifest in a slight change of the functional form.  However, the existence of inertial effects are far more clearly manifest in $\chi'$ due to the presence of a sign change (which as mentioned above is absent in the Debye-like Ryzhkin expression).   Its zero crossing is found at $\sqrt{\frac{\gamma}{\tau}}$.  Phase coherent measures of the spectral response as performed in experiments like THz spectroscopy are essential to detect such a feature \cite{Pan14,Pan15}.  The magnetic conductivity $\kappa$ exhibits both $1/\tau$ and $\gamma$ frequency scales prominently.   The imaginary component of $\kappa$ exhibits a negative extremum at $1/\tau$.   $\gamma$ is seen as the frequency where $\kappa' > \kappa''$ and   $\kappa''$ exhibits a maximum.

The lower left panel of Fig. 1 demonstrates that the bounding behavior of $\chi_m$ is a power law in $1/(i \omega)$, where the power indicates what term in the classical equation of motion is dominant at a particular frequency.   This is because there are three ``response" terms in the equation of motion, which leads to the three terms in the denominator of the susceptibility, which can be written as increasing powers in $- i \omega$.  In different frequency regimes, different terms on the bottom of Eq. \ref{SusceptibilityInertia2} dominate.  Distinct frequency regions of $\frac{1 }{(i \omega)^0}$, $\frac{1 }{(i \omega)^1}$, and $\frac{1 }{(i \omega)^2}$ dependencies are therefore apparent in the plots of  Fig. 1.

\begin{figure}
\includegraphics[width=8cm]{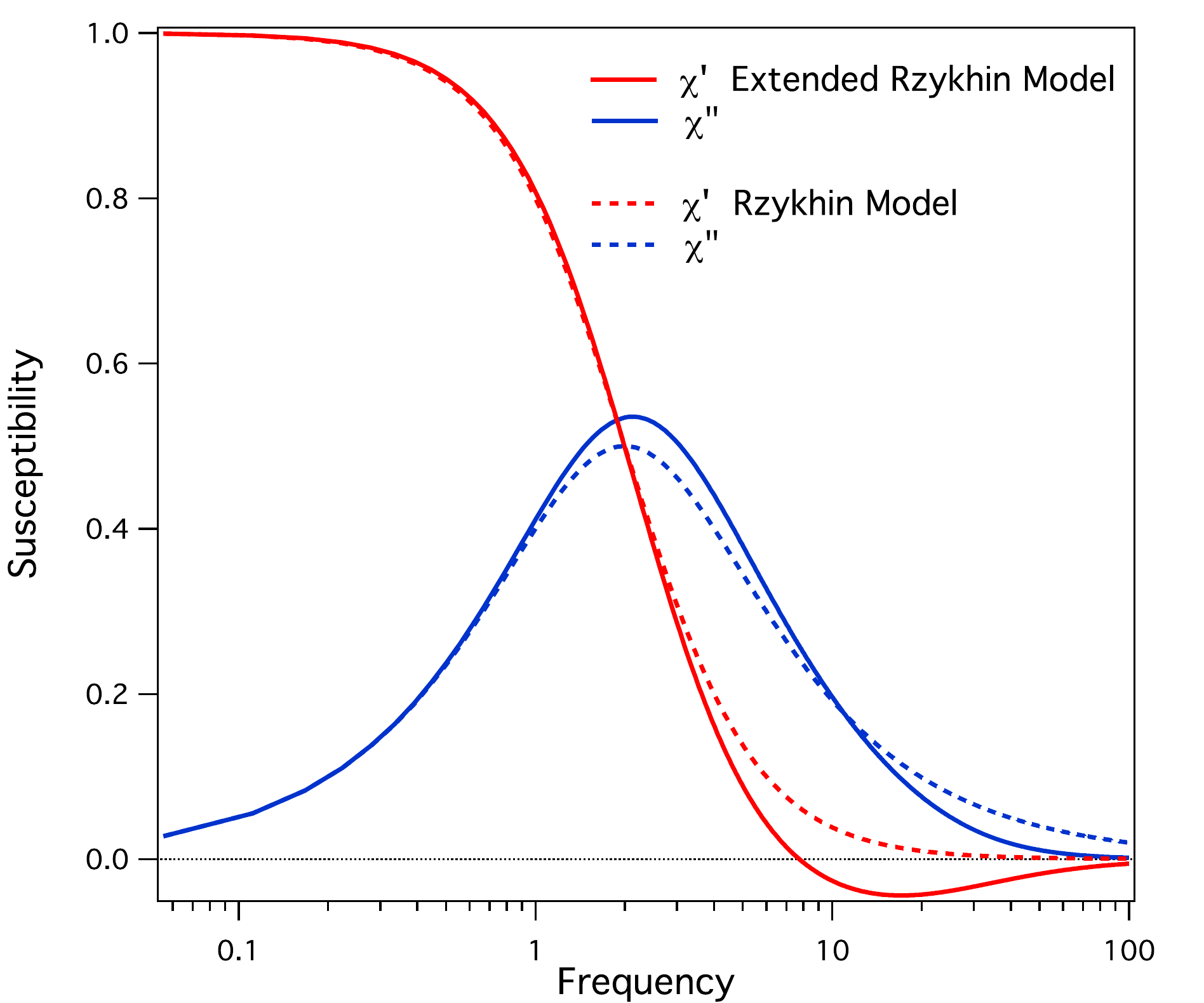}
\centering
\label{SI4} 
\caption{A comparison of the susceptibilities calculated both without and with the inertial term (e.g. Ryzhkin and extended Ryzhkin model) using model parameters $1/ \tau = 2$ and $\gamma = 30$.  }
\end{figure}


\section{Discussion}

From the above discussion on the first-moment sum rule, we learn that any response function must fall off faster than conventional Debye-like relaxation at high enough frequency e.g. other non-relaxational physics must intervene.  In the context of electric charge models, this higher frequency scale is found naturally in inertial effects, but other possibilities may be realizable.  It is important to point out that although some such effect must ultimately intervene, they will not be relevant to the low energy physics if the characteristic frequency scale is high enough.  In the context of the above model, for $\gamma > 1/\tau$, the peak in $\chi"$ is found at $1/\tau$, and has a width of order a few times $1/\tau$.   However the zero crossing of $\chi'$ is found at $\sqrt{\frac{\gamma}{\tau}}$ and so $\gamma$ can in principle be arbitrarily high making inertia irrelevant to the low frequency physics of $\chi_m$, but allowing the sum rule to be formally satisfied.

A few additional comments are appropriate about the applicability of our model for $\chi_m$ for the present situation.  Our present approach was to first define macroscopic variables ($\mathbf{J}, \mathbf{M}$ etc.) in a hydrodynamic-esque fashion and then combine them in a fashion that was consistent with both entropic restoring forces and inertial effects.  The perceptive reader may make the reasonable objection that the long time averaging, which is necessary to define an entropic force is inconsistent with the short time scales on which inertial effects are relevant.  A more rigorous approach would have been to \textit{first} write down a Langevin style equation that includes inertia and \textit{then} do the averaging.  In contrast our approach was one where the order of these operations was essentially reversed.  In general it is not clear if the steps of including the dynamics as such and then averaging or vice versa commute.   Generally, they do not.


Of course, this is not an issue unique to the present case.  Such considerations are ubiquitous when a high frequency response is considered in the context of entropic forces (e.g. high frequency phonon propagation, polymer relaxation etc.).  As discussed above, related issues arise in the inclusion of inertial effects in the context of dielectric relaxation.   A rigorous Langevin treatment of an assembly of non-interacting fixed axis rotators with finite moments of inertia gives a susceptibility that has the complex form of a infinite set of continued fractions \cite{Coffey04}.  Similar expressions can be obtained by different methods \cite{Gross55,Sack57}.  However, for small inertia, this expression can be approximated by the first convergent, and is known as the Rocard equation, after Rocard first derived it in 1933 \cite{Roccard33a} when adding inertial effects to the Debye theory.   The Rocard equation is equivalent to the expression derived above.   This example shows that there are cases and limits in which the ordering in the treatment of inertia and ensemble averaging does not matter.  Unfortunately, it is not known at this time, to what extent these effects matter or how to perform a Langevin style analysis for dynamics in the spin ices.   This is obviously a very open area for theoretical inquiry.

\section{Conclusions}

We can see that based on simple considerations a hydrodynamic approach can be given for the transport of magnetic charge in quantum spin ice systems.   A number of recent results support the possibility of propagating magnetic charge in such systems.  Time-domain THz experiments on Yb$_2$Ti$_2$O$_7$ have given evidence for a sign change in the susceptibility despite very broad spectral features \cite{Pan15}.  As emphasized above, the classical Ryzkhin model is incapable of reproducing this sign change that is naturally accounted for by the inclusion of inertial effects.  This demonstrates a \textit{hidden coherence} in the system despite the broad features.  Analysis of this data gives through the sum rule of Eq. \ref{KappaSumRule} an effective mass very similar to what is expected based on the simplest model of charges hopping on a tight binding lattice with known exchange parameters.   Also in Yb$_2$Ti$_2$O$_7$, Tokiwa et al. \cite{Tokiwa16a} found evidence for long distance propagating particles that carry heat.   Their behavior is consistent with propagating monopoles, although it is not clear that the long length scales they extract are consistent with the reasonable short scattering times extracted from the THz experiments.   Numerically Wan et al. \cite{Wan16a} found in a minimal model for dynamics of monopoles in quantum spin ice that the monopoles propagate as massive quasiparticle at low energy despite their strong coupling to the spin background at the microscopic energy scale.   Irrespective of whether Yb$_2$Ti$_2$O$_7$ is in parameter range that can realize the spin-ice phenomenology, the ideas outlined here should have general applicability in detecting the mass of systems with magnetic charge.

This work was supported under the auspices of the ``Institute for Quantum Matter" DOE DE-FG02-08ER46544.   I would like to thank my collaborators on various related parts of this work including N. J. Laurita, Kate A. Ross, Bruce D. Gaulin and especially LiDong Pan.

\end{document}